\documentstyle[12pt]{article}
\begin{document}
\title {Layered superconductors with long-range Josephson couplings}
\author {by\\Ma{\l}gorzata Sztyren
\thanks{Department of Mathematics and Information Science,
Warsaw University of Technology, Pl. Politechniki 1, PL-00-661 Warsaw
E--mail: emes@mech.pw.edu.pl}}
\date \today

\input{epsf.tex}
\edef\undtranscode{\the\catcode`\_} \catcode`\_11
\newbox\box_tmp 
\newdimen\dim_tmp 
\def\jump_setbox{\aftergroup\after_setbox}
%
%
\def\resize
    #1
    #2
    #3
    #4
    {%
    \dim_r#2\relax \dim_x#3\relax \dim_t#4\relax
    \dim_tmp=\dim_r \divide\dim_tmp\dim_t
    \dim_y=\dim_x \multiply\dim_y\dim_tmp
    \multiply\dim_tmp\dim_t \advance\dim_r-\dim_tmp
    \dim_tmp=\dim_x
    \loop \advance\dim_r\dim_r \divide\dim_tmp 2
    \ifnum\dim_tmp>0
      \ifnum\dim_r<\dim_t\else
        \advance\dim_r-\dim_t \advance\dim_y\dim_tmp \fi
    \repeat
    #1\dim_y\relax
    }
\newdimen\dim_x    
\newdimen\dim_y    
\newdimen\dim_t    
\newdimen\dim_r    
\def\perc_scale#1#2{
  \def\after_setbox{%
    \hbox\bgroup
    \dim_tmp\wd\box_tmp \divide\dim_tmp100 \wd\box_tmp#1\dim_tmp
    \dim_tmp\ht\box_tmp \divide\dim_tmp100 \ht\box_tmp#2\dim_tmp
    \dim_tmp\dp\box_tmp \divide\dim_tmp100 \dp\box_tmp#2\dim_tmp
    \box\box_tmp 
\egroup}%
    \afterassignment\jump_setbox\setbox\box_tmp =
}%
{\catcode`\p12 \catcode`\t12 \gdef\PT_{pt}}
\def\hull_num{\expandafter\hull_num_}
\expandafter\def\expandafter\hull_num_\expandafter#\expandafter1\PT_{#1}
\def\find_scale#1#2{
  \def\after_setbox{%
    \resize\dim_tmp{100pt}{#1}{#2\box_tmp}%
    \xdef\lastscale{\hull_num\the\dim_tmp}\extra_complete}%
  \afterassignment\jump_setbox\setbox\box_tmp =
}
\def\scaleto#1#2#3#4{
  \def\extra_complete{\perc_scale{#3}{#4}\hbox{\box\box_tmp}}%
  \find_scale{#1}#2}
\let\xyscale\perc_scale
\def\zscale#1{\xyscale{#1}{#1}}
\def\yxscale#1#2{\xyscale{#2}{#1}}
\def\xscale#1{\xyscale{#1}{100}}
\def\yscale#1{\xyscale{100}{#1}}
\def\xyscaleto#1{\scaleto{#1}\wd\lastscale\lastscale}
\def\yxscaleto#1{\scaleto{#1}\ht\lastscale\lastscale}
\def\xscaleto#1{\scaleto{#1}\wd\lastscale{100}}
\def\yscaleto#1{\scaleto{#1}\ht{100}\lastscale}
\def\slant#1{
  \hbox\bgroup
  \def\after_setbox{%
    \box\box_tmp 
\egroup}%
  \afterassignment\jump_setbox\setbox\box_tmp =
}%
\def\rotate#1{
  \hbox\bgroup
  \def\after_setbox{%
    \setbox\box_tmp\hbox{\box\box_tmp}
    \wd\box_tmp 0pt \ht\box_tmp 0pt \dp\box_tmp 0pt
    \box\box_tmp
    \egroup}%
  \afterassignment\jump_setbox\setbox\box_tmp =
}%
\newdimen\box_tmp_dim_a
\newdimen\box_tmp_dim_b
\newdimen\box_tmp_dim_c
\def\plus_{+}
\def\minus_{-}
\def\revolvedir#1{
  \hbox\bgroup
   \def\param_{#1}%
   \ifx\param_\plus_ \else \ifx\param_\minus_
   \else
     \errhelp{I would rather suggest to stop immediately.}%
     \errmessage{Argument to \noexpand\revolvedir should be either + or -}%
   \fi\fi
  \def\after_setbox{%
    \box_tmp_dim_a\wd\box_tmp
    \setbox\box_tmp\hbox{%
     \ifx\param_\plus_\kern-\box_tmp_dim_a\fi
     \box\box_tmp
     \ifx\param_\plus_\kern\box_tmp_dim_a\fi}%
    \box_tmp_dim_a\ht\box_tmp \advance\box_tmp_dim_a\dp\box_tmp
    \box_tmp_dim_b\ht\box_tmp \box_tmp_dim_c\dp\box_tmp
    \dp\box_tmp0pt \ht\box_tmp\wd\box_tmp \wd\box_tmp\box_tmp_dim_a
    \kern \ifx\param_\plus_ \box_tmp_dim_c \else \box_tmp_dim_b \fi
    \box\box_tmp
    \kern -\ifx\param_\plus_ \box_tmp_dim_c \else \box_tmp_dim_b \fi
    \egroup}%
  \afterassignment\jump_setbox\setbox\box_tmp =
}%
\def\revolve{\revolvedir-}
\def\xflip{
  \hbox\bgroup
  \def\after_setbox{%
    \box_tmp_dim_a.5\wd\box_tmp
   \setbox\box_tmp
     \hbox{\kern-\box_tmp_dim_a \box\box_tmp \kern\box_tmp_dim_a}%
   \kern\box_tmp_dim_a
    \box\box_tmp
    \kern-\box_tmp_dim_a
    \egroup}%
  \afterassignment\jump_setbox\setbox\box_tmp =
}%
\def\yflip{
  \hbox\bgroup
  \def\after_setbox{%
    \box_tmp_dim_a\ht\box_tmp \box_tmp_dim_b\dp\box_tmp
    \box_tmp_dim_c\box_tmp_dim_a \advance\box_tmp_dim_c\box_tmp_dim_b
    \box_tmp_dim_c.5\box_tmp_dim_c
   \setbox\box_tmp\hbox{\vbox{%
     \kern\box_tmp_dim_c\box\box_tmp\kern-\box_tmp_dim_c}}%
   \advance\box_tmp_dim_c-\box_tmp_dim_b
   \setbox\box_tmp\hbox{%
     \lower\box_tmp_dim_c\box\box_tmp
}%
    \ht\box_tmp\box_tmp_dim_a \dp\box_tmp\box_tmp_dim_b
    \box\box_tmp
    \egroup}%
  \afterassignment\jump_setbox\setbox\box_tmp =
}%
\catcode`\_\undtranscode

\special{ps:}

\maketitle
\newcommand {\eqn}[1]{\begin{equation}#1\end{equation}}
\newcommand{\equln}[2]{\begin{equation} {#1} \label{#2}\end{equation}}
\newcommand{\includegraphics}[1]{\zscale{100}\hbox{\epsffile{#1}}}


\begin{abstract}
The present paper is an extension of {\em cond-mat/0312673}.\\
The construction of a hybrid discrete-continuous model of layered
superconductors is presented. The model bases on the classic
Lawrence-Doniach scenario with admitting, however, long-range
interactions between atomic planes. Moreover, apart from Josephson
couplings they involve the proximity effects. The range of
interactions can, in principle, be arbitrary large. The solutions
corresponding to the range K=2 are found. The mechanism of 
enhancement of superconductivity caused by the proximity effect
and the presence of higher Josephson couplings is shown. The
physical meaning of coupling constants, with particular
attention paid to their sign, is discussed. For the case K=2
the interpretation in terms of microscopic interactions between
Cooper pairs in different planes, as well as the relation to
experimentally measurable quantities, such as the out-of plane
effective mass and bandwith, is given.
\end{abstract}                  


\maketitle

\section{Introduction}

   Most of the high-temperature superconductors like e.g. YBCO
or BiSCCO have a layered structure. Such a strongly
anisotropic situation results in the fact that the material
properties and behaviour of the fields in the direction
(say \(z\) axis)
orthogonal to the layers are totally different from the behaviour
in directions parallel to the layers. The simplest effective
phenomenology of such systems is given by the 3D anisotropic
Ginzburg-Landau model \cite{Blatter+:94,Abrikosov:book,
Rogula:99}.

   However, the abilities of the 3D continuum to describe
discrete array of atomic layers are limited: for a layered
superconductor such a description can be justified only
when \(\xi_c>>s\), where \(s\) is the (typical) interlayer spacing.
The idea originally proposed by Lawrence 
and Doniach (LD) \cite{Lawrence+Doniach:71} is to consider the
layered superconductor as an array of superconducting planes
coupled by Josephson currents \cite{Lawrence+Doniach:71} flowing
perpendicularly between adjacent layers. The
planes themselves are described in terms of the 2D Ginzburg-Landau
phenomenology. Under such circumstances the GL order parameter
\(\psi\) appears as a function of two
continuous variables (say \(x\) and \(y\)) and one discrete variable
\(n\) -- the index of the layer.
The corresponding form of the free-energy
functional was proposed in \cite{Lawrence+Doniach:71}. The model
turned out very successful, and was subsequently modified and enriched
in many papers and textbooks
\cite{ Theo:88, Theo:93, Laza+:93, Ketterson+Song}.
A number of solutions for LD model are given in \cite{Krasnov:2001}
or \cite{Kuple:2001}.

  As compared with the 3D anisotropic GL, the LD model can be
considered the other extreme case, justified when \(\xi_c{{\leq}}s\).
However, both the models together do not cover the entire range of
the coherence lengths \(\xi_c\) in relation to the spacing \(s\):
there is a gap between the models corresponding to coherence
lengts \(\xi_c\) which 
are too small for GL and too big for LD. 

\indent 
The higher grade hybrid model, proposed in the present paper,
patches the above gap by
admitting direct couplings not only between adjacent, but also
between more distant atomic planes. Such couplings can be
particularly expected due to strong interplanar electronic
correlations.
The low-temperature correlation lengths of many superconducting
materials, such as e.g.
Bi\(_2\)Sr\(_2\)CaCu\(_2\)O\(_8\), with \(\xi_c=28\){\AA}, or
Tl\(_2\)Ba\(_2\)CaCu\(_2\)O\(_8\), with \(\xi_c=23\){\AA},
do not justify the restriction of interplanar couplings to
nearest neighbours only. Since in both cases the interplanar spacing
\(s\) is about 15{\AA}, the ratio \(\xi_c/s\) ranges, roughly speaking,
from 1.5 to 2, and this suggests the need for interplanar couplings
including nearest and next-nearest planes at least.
The situation here is quite analogous to that in the theory of
crystalline lattices: in general the nearest
neighbour couplings alone are not sufficient for adequate description
of lattice dynamics \cite{Born+Huang}.

The paper is organized as follows. In the next sections we give
a formal presentation of our hybrid model with couplings among more
distant neighbours -- up to a (given, but arbirary) range \(K\),
with emphasis on \(K=2\). Subsequently we determine the
ground state solutions and discuss their stability in function
of admissible coupling parameters. For the sake of brevity,
in other (frequently important) respects we keep our
considerations as close to
classic LD model as possible, avoiding e.g. heterostructures,
non-uniform interplanar spacings, or else gradient coupling
\cite{Theo:93}. Those factors will be discussed elsewhere.

  In Section \ref{enha} we discuss the relation between 2D and 3D
superconductivity in our model.
The presence of interplanar couplings modifies the in-plane
superconductivity. Depending on the coupling constants
the modification can turn out supression or enhancement.
In particular, even overcritical planes 
can, under appropriate couplings, result in subcritical
superconductivity of the whole array.

If the nearest-neighbour coupling dominates, one can pass to \(K=1\).
The resulting hybrid model with \(K=1\) offers, however, more
possibilities than the classic LD; the latter obtains as a specialized
case of this variant.
The relevant questions, including the relation to other existing
models, are addressed in the Section \ref{spec}.
The final Section \ref{disc} is
devoted to discussion of the physical significance of the coupling
constants, with particular attention paid to their signs.

\section{The hybrid model of grade K}\label{hybr}

   Let us regard the layered superconductor
as a one-dimensional chain of atomic planes
with Josephson's bonds between them. Such bonds
will be called J-links. The interplanar distance will be denoted
by \(s\). We assume
the following convention for indexing the planes and links.
If we locate the point \(z=0\) at an atomic plane,
then the \(z\)-coordinate of
any plane, equal \(ns\), may be represented by the integer \(n\),
while the \(z\)-coordinate of the center of any interplanar gap,
equal \(ls\),
by the half-integer \(l\). Choosing the point \(z=0\) at the center
of an interplanar gap -- we index the planes by half-integers and the
gaps by integers.

   Let us consider the free-energy functional \(\cal{F}\) for
a layered superconductor. We shall denote by \(\psi_n\) the order
parameter associated to the layer indexed by the number \(n\). Its
complex conjugate (c.c.) will be denoted by \(\bar{\psi}_n\).
The symbol \(m_{ab}\) and \(m_c\) will denote the in-plane and
out-of-plane effective mass of superconducting current carriers,
respectively.
We start from the free energy functional of the following form 
\begin{equation}
{\cal{F}}={\cal{F}}_0+{\cal{F}}_s+\frac{1}{8\pi}\int {\bf B}^2d^3x.
\label{calf}
\end{equation}
The term \({\cal F}_0\) describes the normal state, while \({\cal F}_s\)
the superconducting one. The supercoducting term is composed
of two parts: 
\begin{equation}
{\cal{F}}_s={\cal{F}}_p+{\cal{F}}_J,\label{fs}
\end{equation}
where the part
\begin{equation}
{\cal{F}}_p=s\sum_n\int dxdyF_n
\end{equation}
describes the contribution of atomic planes, while
\({\cal{F}}_J\) corresponds to interplanar Josephson's bonds.
For any plane indexed by \(n\) the free energy density
\(F_n\) has the 2D Ginzburg-Landau form
(in general, the parameters can depend on \(n\))
\begin{equation}
F_{n}=\frac{\hbar^2}{2m_{ab}}|({\bf D}\psi)_n|^2 
 +\alpha_0|\psi_n|^2+\frac{1}{2}\beta|\psi_n|^4,\label{efn}
\end{equation}
where we have introduced the 2-dimensional continuous operator \({\bf D}\)
(covariant derivative)
\begin{equation}
D_{\rho}=\partial _{\rho}-\frac{ie^*}{\hbar c}A_{\rho},
 \ \ \ \rho=x,y.
\end{equation}

The form (\ref{efn}) of the functional \({\cal{F}}_p\) already 
ensures its invariance with respect to the gauge transformation
\begin{equation}
\left\{ \begin{array}{l}
{\bf A}\rightarrow {\bf A}'={\bf A}+\nabla\Lambda,\\
\\
 \psi_n\rightarrow \psi'_n=\psi_ne^{i\frac{e^*}{\hbar c}\Lambda},\\ 
\\
 \bar{\psi}_n\rightarrow \bar{\psi}'_n=
 \bar{\psi}_ne^{-i\frac{e^*}{\hbar c}\Lambda}\label{gau}\\
\end{array}
\right .
\end{equation}
(2-dimensional {\bf A} and \({\bf \nabla}\) for this case). The standard variational treatment
of the functional \({\cal F}_s\) with
respect to \(A_{\rho},\ \ \rho=x,y\), gives
the standard in-plane components of the supercoducting current
\begin{equation}
j_{\rho}=-\frac{ie^*\hbar}{2m_{ab}}(\bar{\psi}_n\partial_{\rho}\psi_n
     -\psi_n\partial_{\rho}\bar{\psi}_n)-\frac{e^{*2}}{m_{ab}c}A_{\rho}|\psi_n|^2
           ,\ \ \rho=x,y.\label{incur}
\end{equation}

   Let us now construct the term \({\cal{F}}_J\) in the free-energy.
In general it is a functional which can depend on all
\(\psi_n, \bar{\psi}_n\) and
on the vector potential \({\bf A}\).
The simplest gauge invariant expression for the energy of J-link
between \(n\)-th and (\(n+q\))-th planes will have the following form
\begin{equation}
\varepsilon_{qn}=\frac{1}{2}\{\zeta_{qn}
 \bar{\psi}_n\psi_n+\eta_{qn}
 \bar{\psi}_{n+q}\psi_{n+q}-(\gamma_{qn}
 \bar{\psi}_n\psi_{n+q}e^{-ip_{qn}}+c.c.)\},\label{eps1}
\end{equation}
where the exponent \(p_{qn}\) is defined by the formula
\begin{equation}
p_{qn}=\frac{e^*}{\hbar c}\int_{ns}^{(n+q)s}A_zdz.\label{pqn}
\end{equation}
Let us note that for the model invariant with respect to the time
reversal we have \(\gamma_{qn}=\bar{\gamma}_{qn}\).
In general, the coupling parameters \(\zeta,\ \eta,\ \gamma\)
as well as in-plane parameters \(\alpha_0\) and \(\beta\) can be
different for different planes (which is the case for superconductors
composed of various atomic planes). However, in this paper we shall
consider the array of identical planes, hence the parameters will not
depend on the index \(n\). That implies \(\eta_q=\zeta_q\). Hence
instead of (\ref{eps1}) we shall use
\begin{equation}
\varepsilon_{qn}=\frac{1}{2}\{\zeta_q
(|\psi_n|^2+|\psi_{n+q}|^2)
-\gamma_q
 (\bar{\psi}_n\psi_{n+q}e^{-ip_{qn}}+c.c.)\}.\label{epsi}
\end{equation}
Let \(P\) denote the (finite or infinite, but ordered) set of all
indices of planes, and let \(Q=\{1, 2,..., K\}\).
The planes connected with the \(n\)-plane by Josephson
coupling select the following subset of \(Q\)
\begin{equation}
P_n=\{q\epsilon Q:\ (n+q)\epsilon P\}.\label{pn}
\end{equation}
Thus, the gauge invariant functional \({\cal{F}}_J\)
for the hybrid model of grade K has the
following form
\begin{equation}
{\cal{F}}_J=s\sum_{n\epsilon P}\sum _{q\epsilon P_n}
\int dxdy\,\varepsilon_{qn},\label{fj}
\end{equation}
with \(\varepsilon_{qn}\) given by (\ref{epsi}).
The coupling parameters \(\zeta_q\) and \(\gamma_q\) vanish
for \(q>K\). Every J-link is represented in (\ref{fj}) by exactly
one term.

\section{Comparison with anisotropic GL model}\label{compa}

To compare our hybrid model (HM) with the continuum GL model,
we shall consider the infinite medium. In that case the summation
index may be shifted by the integer \(q\).
Moreover \(P_n=Q\).
This implies that
\begin{equation}
F_J=\frac{1}{2}\sum_n\sum_q\{2(\zeta_q-\gamma_q)|\psi_n|^2
 +\gamma_q|\psi_{n+q}e^{-ip_{qn}} -\psi_n|^2\}.\label{FJ}
\end{equation}
For very weak field \(A_z\) and very slow dependence of \(\psi_n\)
on \(n\) we have the correspondence rules which, in the long wave
limit, allow us
to pass from the hybrid to the 3D continuum:
\begin{equation}
\left\{ \begin{array}{l}
\sum_n \rightarrow \frac{1}{s}\int dz,\\
\\
\frac{1}{qs}(\psi_{n+q}-\psi_n) \rightarrow \psi'_n(z),\\
\\
e^{-ip_{qn}} \rightarrow 1-i\frac{e^*}{\hbar c}qsA_z.\\
\end{array}
\right .
\label{rules}
\end{equation}
Applying the rules to the functional (\ref{fs}) with (\ref{efn})
and (\ref{FJ}) one obtains
%
\begin{equation}
{\cal F}_s \rightarrow \int d^3x\{\frac{\hbar^2}{2m_{ab}}
 |{\bf D}\psi|^2
 +\alpha_0|\psi|^2+\frac{1}{2}\beta|\psi|^4
 +\sum_q[(\zeta_q-\gamma_q)|\psi|^2+\frac{1}{2}q^2s^2\gamma_q|D_3\psi|^2]\}
.
\end{equation}
%
Hence, we have the following relation between \(m_c\) -- the effective
mass in \(z-\)direction (in the anisotropic GL model) and the coupling
parameters \(\gamma_q\):
\begin{equation}
\frac{1}{m_c}=\frac{s^2}{\hbar^2}\sum_qq^2\gamma_q.
\label{mass}
\end{equation}
The presence of J-links modifies also the parameter \(\alpha_0\)
to the form:
\begin{equation}
\alpha=\alpha_0+\sum_q(\zeta_q-\gamma_q).\label{alf}
\end{equation}

  The condition of weak dependence of \(\psi_n\) on \(n\),
valid under usual assumption of nearly uniform states,
is crucial for the transition from discrete to continuum picture
in the z-direction. For strongly oscillating dependence, in
particular for alternating ground-state solutions discussed
in the sequel, transition (\ref{rules}) can no longer be established.
Nevertheless, sufficiently weak and sufiiciently smooth
excitations from a given ground state can
again be described by an effective 3D anisotropic GL model,
specific for this ground state. The equations (\ref{mass})
and (\ref{alf})
remain valid only for excitations with respect
to the uniform ground-state.
In the case of phase-oscillatory ground states
\(\stackrel{{\rm o}}{\psi}_n \), making use of the substitution
\begin{equation}
 \psi_n \rightarrow\ \stackrel{{\rm o}}{\psi}_n (1+\psi_n)
\end{equation}
with a suitable reinterpretation of the correspondence rules
(\ref{rules}),
one arrives at the modified equations, which can be obtained
from (\ref{mass}) and (\ref{alf}) through the replacement
\begin{equation}
 \gamma_q \rightarrow \gamma_q \cos(q\theta).
\end{equation}
The requirement of the overall stability of the system is
responsible for the positive sign of the effective mass \(m_c\).

\section{Field equations}\label{field}

By computing the variation of the functional \({\cal F}\) with
respect to \(A_z\) one obtains the Maxwell equation for the
\(z-\)components
of current density and \({\rm curl}\, H\)
\begin{equation}
\frac{1}{c}J(z)=\frac{1}{4\pi}({\rm curl}\, H)_z,
\end{equation}
where 
\begin{equation}
J(z)=\frac{se^*}{2i\hbar}
\sum_{n\epsilon P}\sum_{q\epsilon P_{n}}
 \{\gamma_q\bar{\psi}_n\psi_{n+q}e^{-ip_{qn}}-c.c.\}\chi_{qn}(z),
 \label{cna}
\end{equation}
\(P_n\) is given by (\ref{pn}), and the quantity \(p_{qn}\) by (\ref{pqn}).
The symbol \(\chi_{qn}(z)\) denotes
the characteristic function of the interval \([ns,(n+q)s]\). Let us note
that for any layer \(ns<z<(n+1)s\) the exppression \(J(z)\) does not
depend on the value z; only the ends of the interval are important. To
better see the structure, let us first extend the set \(Q\)
on the negative values
\begin{equation}
\bar{Q}=\{-K,...,-2,-1,1,2,...,K\},
\label{qbar}
\end{equation}
and introduce the symbols
\begin{equation}
\sigma_{qn}=\left\{ \begin{array}{l}
1\ \ {\rm if}\ \ \ (n+q)\ \epsilon P\ ,
\\
0\ \ {\rm otherwise}\ .
\end{array}
\right .\label{sig}
\end{equation}
Then the expression for Josephson current \(J_l\) describing tunneling
across the interplanar gap indexed by \(l\) (half integer if \(P\)
contains integers) will have the form
\begin{equation}
J_l=\frac{se^*}{2i\hbar}\sum_{q\epsilon \bar{Q}}
\sum_{n\epsilon P_{lq}}
 \{\gamma_q\sigma_{qn}\bar{\psi}_n\psi_{n+q}e^{-ip_{qn}}-c.c.\},
 \label{cnb}
\end{equation}
where
\begin{equation}
P_{lq}=\{n\,\epsilon\,P: n<l<n+q\}.
\end{equation}
By computing the variation of the functional
 \({\cal F}_s\) with
respect to \(\bar{\psi}_n\), one obtains the equations

\begin{equation}
  -\frac{\hbar^2}{2m_{ab}}{\bf D}^2\psi_n+\tilde{\alpha}\psi_n+
     \beta|\psi_n|^2\psi_n
    -\frac{1}{2}\sum_{q\epsilon \bar{Q}}
    \gamma_q
    \sigma_{qn}\psi_{n+q}e^{-ip_{qn}}=0
\label{eqpsi}
\end{equation}
where instead of \(\alpha_0\) we have introduced
\begin{equation}
\tilde{\alpha}=\alpha_0+\frac{1}{2}
      \sum_{q\,\epsilon \bar{Q}}\sigma_{qn}\zeta_q\label{tilal}
\end{equation}
which, for finite \(P\), depends of \(n\).

\section{The ground states}\label{ground}

    Let us now consider the plane-uniform states of HM in
the absence of magnetic field. The order parameter is then
independent of the in-plane variables and the net supercurrents
vanish. In detailed calculations we shall focus our attention
on the grade \(K=2\), which seems to be sufficiently illustrative
to grasp the idea on what is going on. The generalization to
grades \(K>2\), although more complicated algebraically,
is straightforward. 
In the region far from the boundary all the coefficients \(\sigma_{qn}=1\).
    For \(K=2\) the condition of vanishing Josephson current
is equivalent to
\begin{equation}
\gamma_1(\bar\psi_n\psi_{n+1}-c.c.)
 +\gamma_2(\bar\psi_n\psi_{n+2}
 +\bar\psi_{n-1}\psi_{n+1}-c.c.)=0,
\end{equation}
and the equations (\ref{eqpsi}) take the form
\begin{equation}
\tilde{\alpha}\psi_n +\beta|\psi_n|^2\psi_n-\frac{1}{2}[\gamma_1
(\psi_{n+1}+\psi_{n-1})+\gamma_2(\psi_{n+2}+\psi_{n-2})]=0.
\label{far}
\end{equation}
We shall look for  solutions with constant amplitude and
difference of phase between adjacent atomic planes, so we use the
ansatz
\begin{equation}
\psi_n=Ce^{in\theta}.
\end{equation}
The result is the equation
\begin{equation}
\tilde{\alpha}+\beta C^2-\gamma_1 \cos\theta-\gamma_2\cos 2\theta=0,
\label{upsi}
\end{equation}
with the condition
\begin{equation}
\gamma_1 \sin\theta+2\gamma_2 \sin 2\theta=0.\label{teta}
\end{equation}
Solving (\ref{teta}) with respect to \(\theta\), we obtain 3 variants:
(a1) \(\theta=0\), (a2) \(\theta=\pi\), and
(a3) \(\cos\theta=-\gamma_1\)/\(4\gamma_2\). The solution \(C\) to (\ref{upsi})
has the form \(C^2=-\alpha^*\)/\(\beta\), where \(\alpha^*\) depends
on the variant.
The variant (a1) implies the uniform solution to (\ref{far}):
\begin{equation}
\psi_n=C,\label{const}
\end{equation}
with
\begin{equation}
\alpha^*=\alpha_0 + \zeta_1 + \zeta_2
-\gamma_1 - \gamma_2,\label{al1}
\end{equation}
what is the case of equation (\ref{alf}) for grade K=2.
In the variant (a2) the solution to (\ref{far}) has the alternating form
\begin{equation}
\psi_n=\pm C,\label{alt}
\end{equation}
with  
\begin{equation}
\alpha^*=\alpha_0 + \zeta_1 + \zeta_2
+\gamma_1 - \gamma_2.\label{al2}
\end{equation}
Finally, in the variant (a3), the solution exists if the parameters
\(\gamma_1\) and \(\gamma_2\) fulfill the relation
\begin{equation}
|\gamma_1|\,{\leq}4|\,\gamma_2|.\label{gamy}
\end{equation}
Then the parameter \(\alpha^*\) is connected with the coupling
constants by the formula:
\begin{equation}
\alpha^*=\alpha_0+\zeta_1+\zeta_2+\gamma_2(1+\frac{\gamma_1^2}
  {8\gamma_2^2}).\label{al3}
\end{equation}
There are two independent solutions
\begin{equation}
\psi_n=Ce^{\pm i\theta},\ \ 
  \theta={\rm \arccos}(-\frac{\gamma_1}{4\gamma_2}).\label{modu}
\end{equation}
They will be referred to as the phase modulated states.
The solutions degenerate at the extremities
\(|\gamma_1|\,=4|\,\gamma_2|\).

    So far we confined our discussion to the existence of
solutions which could serve as candidates for the ground state.
However, unstable candidates, even uniform ones, have to be
rejected. Note that alternating states were already found in the
literature on different grounds -- phenomenological \cite{Theo:88}
as well as microscopic ones \cite{Laza+:93}.
The question of stability of the solutions
will be addressed in the next section.

    Let us note that the condition \(\sin\theta=0\) admits
the solutions (\ref{const}) and (\ref{alt}) for
hybrid model of any grade K. Contrary to that, the analogues of
the conditions (a3) and (\ref{gamy}) can deliver, depending on the grade
and on the coupling parameters, any number from \(0\) to \(2(K-1)\)
modulated solutions.

\section{Stability}\label{stabil}

To examine the stability of the solutions presented in
Section \ref{ground}, we shall analyze the Hessian matrix of the 
free energy \({\cal F}_s\) describing small deviations from the ground state.
The problem reduces to examining the function
\begin{equation}
{\cal E}(C,\theta)=\tilde{\alpha}C^2+\frac{1}{2}\beta C^4
 -C^2(\gamma_1 \cos \theta+\gamma_2 \cos 2\theta).\label{stab}
\end{equation}
The solutions found in the previous section are stationary points
of \({\cal E}\).
If \(\gamma_2=0\), then the stability of the solutions depends on
the sign of \(\gamma_1\). If \(\gamma_2\, \neq\, 0\) then 
the sign of the respective eigenvalue
depends on the values of the parameters \(\gamma_1\)
and \(\gamma_2\) . For \(\theta\) different from \(0\) and \({\pi}\)
the dependence has form according to the function
\begin{equation}
f(\gamma_1,\gamma_2)=\gamma_2(\gamma_1+4\gamma_2)(\gamma_1-4\gamma_2)
.\label{deter}
\end{equation}
The straight lines \(\gamma_1+4\gamma_2=0\) and \(\gamma_1-4\gamma_2=0\)
divide the plane \(\gamma_1,\gamma_2\) into 4 regions (see Fig. \ref{figa}).\\
\indent As explained above, both the uniform and the
alternating solutions always exist. However, in the region
\( \gamma_1 >0,\ \gamma_1+4\gamma_2>0\),
only the uniform solution (\ref{const}) is stable,
while in the region
\( \gamma_1 <0,\ \gamma_1-4\gamma_2<0\),
only the alternating solution (\ref{alt}) is stable. The region
(S): \( \gamma_2 <0,\ 4\gamma_2<\gamma_1<-4\gamma_2\),
excludes the stability of both the uniform and the alternating
solutions
but, in contrast to that, ensures the existence and stability of
the modulated solutions (\ref{modu}).
In the region (N): \( \gamma_2 <0,\ -4\gamma_2<\gamma_1<4\gamma_2\),
the modulated solutions exist but are unstable. 

   Let us note that in the regions (NW) and (W) of
stability of the alternating solution (\ref{alt}) one can apply
the construction of bi-continuum solution presented in
\cite{Sztyren:2002}. Note also that, irrespectively of the values
and signs of \(\gamma_1,\,\gamma_2\), a stable ground state
solution always exists.

\section{Enhancement of the superconductivity}\label{enha}

The asociation of the formulae (\ref{al1}), (\ref{al2}) and (\ref{al3})
with the regions of stability of the ground state shows that,
for suitable relations between the coupling constants
\(\gamma_1\) and \(\gamma_2\), one can make the parameter
\(\alpha^*\) more negative than \(\alpha_0\). In consequence,
the 3D superconductivity can turn out enhanced with respect to
the 2D superconductivity in the atomic planes. Such a possibility
has been indicated in the literature \cite{Anderson:92}. The presence
of coupling constants \(\zeta_q\) allows to describe the proximity
effect between atomic planes.

  The corresponding shift of the parameter \(\alpha\) depends on
both \(\zeta\)'s and \(\gamma\)'s. While the contribution from
\(\zeta\)'s reduces always to simple renormalization of \(\alpha_0\)
by the same additive term
\(\zeta_1+\zeta_2\), the contribution from \(\gamma\)'s
in (\ref{al1}), (\ref{al2}) and (\ref{al3}) varies.
To expose the dependence on \(\gamma\)'s, we shall count
the enhancement with respect
to the origin located at \(\tilde{\alpha}\) given by (\ref{tilal}). 
It is convenient to introduce the polar coordinates \(\gamma\)
and \(\varphi\) in the plane \(\gamma_1,\ \gamma_2\):
\begin{equation}
\gamma_1=\gamma\cos\varphi,\ \gamma_2=\gamma\sin\varphi,
\end{equation}
and the notation
\begin{equation}
\varphi_0=\arctan(\frac{1}{4}).
\end{equation}
The quantity 
\(\Delta\alpha=\alpha^*-\tilde{\alpha}\) as a function of the
coupling angle \(\varphi\) is plotted in Fig. \ref{figb}
(the numerical values are computed for
\(\gamma=1\)).\\
  In the uniform state (\ref{al1}) we have
\begin{equation}
\Delta\alpha=\sqrt{2}\gamma\sin(\varphi-\frac{3}{4}\pi), 
 \ -\varphi_0\,{\leq}\varphi\,{\leq}\pi-\varphi_0.
\end{equation}
The minimum \(\Delta\alpha\) (hence the maximum enhancement)
is reached at \(\varphi=\pi\)/\(4\) and equals \(-\sqrt{2}\gamma\).

   In the alternating state (\ref{al2}), in turn, one obtains
\begin{equation}
\Delta\alpha=\sqrt{2}\gamma\sin(\varphi+\frac{3}{4}\pi),
 \ \ \varphi_0\,{\leq}\varphi\,{\leq}\,\pi+\varphi_0
\end{equation}
with the minimum value \(-\sqrt{2}\gamma\) reached at 
\(\varphi=\frac{3}{4}\pi\).

   The enhancement for the phase modulated state (\ref{al3})
is represented by
\begin{equation}
\Delta\alpha=\gamma(1+\frac{1}{8}{\rm ctg}^2\varphi)\sin\varphi,
 \ -\pi+\varphi_0\,{\leq}\,\varphi \,{\leq}\,-\varphi_0.
\end{equation}
In this case the minimum equals \(-\gamma\) and is reached at \(\varphi=-\pi\)/\(2\).
Hence, the maximum enhancement in this case is smaller than
in the uniform and alternating states.

   Let us note that, in the enhancement mechanism discussed above,
the 2D superconductivity of the planes is not a prerequsite for
the 3D superconductivity of the array of the planes. In fact, one can
obtain the negative \(\alpha^*\) starting from
positive \(\alpha_0\). This is in concordance with ideas expressed
in Anderson's discussion of his Dogma V in \cite{Anderson:97:book}.

\section{Special cases}\label{spec}

The HM with K=1 has in general two
coupling parameters \(\zeta\) and \(\gamma\). If they are equal to
one another, one obtains either the Lawrence-Doniach
\cite{Lawrence+Doniach:71}
or Theodorakis \cite{Theo:88} model, depending on the sign
of \(\gamma\). The formula for 
the effective mass
in z-direction simplifies to
\begin{equation}
\frac{1}{m_c}=\frac{s^2}{\hbar^2}|\gamma|.
\end{equation}
   Due to \(\gamma>0\) the LD model has only one
stable  solution, namely the uniform ground state.
The effective parameter \(\alpha=\alpha_0\), and 
the interplanar coupling
gives neither enhancement nor suppression of the
critical temperature. Although the Josephson current coupling
places the model on the enhancement side, the efect is
precisely annuled by \(\zeta=\gamma\).

  Where the negative \(\gamma\) is concerned, the
ground state solution is also unique. Contrary, however, to
the LD case, this solution is alternating, the proximity effect
is present, and the effective
parameter \(\alpha=\alpha_0-2|\gamma|\) results in
enhancement of the supeconductivity.

\section{Discussion}\label{disc}
In the present section we shall discuss the physical meaning of the
coupling constants \(\gamma_1,\ \gamma_2\) with special attention
paid to their signs. 

   {\bf 1.} First of all it is expedient to
note a certain symmetry between
hybrid systems with negative and positive \(\gamma_1\). To that end
consider the transformation
\begin{equation}
\left\{ \begin{array}{l}
\psi'_n=(-1)^n\psi_n,\\
{\bf A}'={\bf A},\\
\gamma'_q=(-1)^q\gamma_q,\\
\end{array}
\right .
\label{trans}
\end{equation}
with all the remaining quantities kept fixed. This transformation
switches between systems with all the constants \(\gamma_q\)
transformed according to the parity of \(q\): the constants remain
identical
for even \(q\)'s and change sign for odd \(q\)'s; in particular
\(\gamma'_1=-\gamma_1\) and \(\gamma'_2=\gamma_2\). At the same time
the order parameter \(\psi_n\) is transformed by the
sign-alternating factor; in particular the states of uniform
order parameter are
transformed into alternating ones, and {\em vice versa}.

  An inspection of the 
equation (\ref{eps1}) shows that the energy of the Josephson's coupling
is invariant with respect to transformation (\ref{trans}). This invariance
extends to the whole free energy functional (\ref{calf}). Note, in
particular, the correspondig symmetry in Fig. 1 and Fig. 2.
Two hybrid systems related by the transformation (\ref{trans}) have
identical ground state energies, excitation energy spectra, and all the
thermodynamical properties, including \(T_c\).

   {\bf 2.} The second
coupling constant \(\gamma_2\) represents the microscopic interactions
between Cooper pairs located in next-nearest layers.
Similarly as in the case of \(\gamma_1\), corresponding to \(q_3\)
in \cite{Laza+:93}, the sign of \(\gamma_2\)
reflects the attractiveness or repulsiveness of those interactions,
both signs being equally well admissible on physical grounds.
Contrary, however, to \(\gamma_1\), the sign of \(\gamma_2\) 
can not be relativized by the
transformation (\ref{trans}). To discuss this situation let us first
select one of the admissible ground states determined
in Section \ref{ground}, and then consider the related excited states
resulting from \(z\)-dependent variations of the order parameter
with respect to the selected groumd state.

   {\bf 3.} For simplicity we start here from
the case of sufficiently
weak coupling between next-nearest layers, so that
\(|\gamma_2|<|\gamma_1|/4\), what corresponds to the
western and eastern sectors in Fig. 1, with the alternating and
uniform ground states, respectively. The
strong coupling sectors
will be commented in the sequel.

  Understood
as the excess energy over the ground state and considered as
a fuction 
of the wave vector,
the excitation energy equals
\begin{equation}
\epsilon_k=\frac{\hbar^2}{2m_{ab}}(k_x^2+k_y^2)+
|\gamma_1|(1-\cos{k_zs})+\gamma_2(1-\cos{2k_zs}).
\label{ek}
\end{equation}
Note that the contribution from \(\gamma_1\) is always positive.

  In the present work it will be sufficient to consider only excitations
with \(k_x=k_y=0\). They are characterized by the effective
mass
\begin{equation}
m_c=\frac{\hbar^2}{s^2}(|\gamma_1|+4\gamma_2)^{-1}
\label{efm}
\end{equation}
in the vicinity of the ground state \(k=0\),
and by the bandwidth
\begin{equation}
W=2|\gamma_1|; 
\label{bandWE}
\end{equation}
the latter furnishes
an alternative interpretation of \(\gamma_1\) in this case.
Provided that the
bandwidth \(W\) and the effective mass \(m_c\) (as well as the
interlayer distance \(s\)) are known from
measurements, the coupling constant \(\gamma_2\) can be determined
from the equation
\begin{equation}
\gamma_2=\frac{1}{8}(\frac{4}{M} - W)
\label{gad}
\end{equation}
where 
\begin{equation}
M=\frac{2s^2m_c}{\hbar^2}
\label{M}
\end{equation}
represents the effective mass, scaled conveniently
for the sequel. The inequalities
\begin{equation}
2\ {{\leq}}\ MW\ {{\leq}}\ \infty
\label{wein}
\end{equation}
hold in the western and eastern sectors.

   The quantity \(W_{LD}=4/M\)
represents the bandwidth for
\(\gamma_2=0\), which corresponds to the LD model. This bandwidth is
entirely determined by the spacing \(s\) and the effective mass
\(m_c\) alone, what is one
of artifacts due to oversimplification of the model resulting
from \(\gamma_2=0\). In our model the effective mass and the
bandwidth are independent measurable parameters. According
to equation (\ref{gad}), the sign of \(\gamma_2\) is directly
determined by the relation between the measured bandwidth \(W\)
and the parameter \(W_{LD}\) calculated from the effective mass:
when \(W<W_{LD}\), the constant \(\gamma_2\) is positive, and
when \(W>W_{LD}\), it is negative.

  {\bf 4}. Now we shall briefly discuss the case of strong
next-nearest coupling,
defined by the relation (\ref{gamy}), represented by the remaining
two sectors in Fig. 1. The dispersion
(\ref{ek}) is no longer monotonic,
and in place of the simple
expression (\ref{bandWE}) for the bandwidth we have
\begin{equation}
W=|\gamma_1|+2|\gamma_2|+\frac{\gamma_1^2}{8|\gamma_2|},
\label{bandNS}
\end{equation}
the formula common for both sectors.
However, the ground-state effective masses in the northern
(\(\gamma_2>0)\) and the southern (\(\gamma_2<0)\) sector differ
from one another. 

    In the northern sector we have, according to the results of
Section \ref{ground}, two stable solutions with generically different
energies: the ground state and the metastable state. For
the ground state effective mass the equation (\ref{efm}) is still
valid.
Combined together, the equations (\ref{bandNS}) and (\ref{efm}) allow
to establish the inequalities
\begin{equation}
1\ {{\leq}}\ MW\ {{\leq}}\ 2
\label{nin}
\end{equation}
which restrict the admissible range of \((m_c, W)\) in the northern sector.
Provided that these restrictions are satisfied, the equations
(\ref{bandNS}) and (\ref{efm}) furnish the unique solution
for the coupling constants
\begin{equation}
|\gamma_1|=\frac{2}{M}(1-\frac{1}{MW}),
\label{ngama}
\end{equation}
\begin{equation}
\gamma_2=\frac{1}{2M^2W}.
\label{ngamb}
\end{equation}

   The southern sector is characterized by two symmetric
phase-modulated
ground states. The corresponding effective mass is given by
\begin{equation}
m_c= \frac{\hbar^2}{s^2} \frac{4|\gamma_2|}{16\gamma_2^2-\gamma_1^2}
\label{smass}
\end{equation}
and in place of the inequalities (\ref{nin}) we obtain
\begin{equation}
1\ {{\leq}}\ MW < {\infty}.
\label{sin}
\end{equation}
Finally, under the above restriction, the coupling
constants are uniquely determined from \((M, W)\) as
\begin{equation}
|\gamma_1|=\frac{1}{2}(W-\frac{1}{M^2W}),
\label{sgama}
\end{equation}
\begin{equation}
\gamma_2=-\frac{W}{8}(1+\frac{1}{MW})^2.
\label{sgamb}
\end{equation}

   {\bf 5.} For a known sector, making use of the relevant pair of
equations (\ref{bandWE})-(\ref{gad})
or (\ref{ngama})-(\ref{ngamb})
or else (\ref{sgama})-(\ref{sgamb}), one can uniquely
determine \(|\gamma_1|\) and \(\gamma_2\) from
experimental values of the effective mass \(M\) and
the bandwidth \(W\). However, identification of the sector
with the aid of inequalities (\ref{wein}), (\ref{nin}), and
(\ref{sin}) is
possible only partially: eg. west, east and northern sectors can
be mixed with the southern one, while \(MW=2\) constitutes the border
between west and east sectors on one side, and the northern sector
on the other side.
Nevertheless, more detailed knowledge concerning
identification of the sectors
can be gained from closer examination
of the excitation spectra. In particular, a feature which
distiguishes sectors N+S from W+E is the presence
of a strong in-band singularity of the density of states
\(N(\epsilon_k)\).

   {\bf 6.} In consequence of the symmetry (\ref{trans}),
one can not experimentally
discriminate between sectors W and E by measurements
based on the ground state or excitation energies, or else on any
physical quantities determined
by the energy spectra. Nevertheless, such a discrimination should
be possible with the aid of suitably chosen measurable physical
quantities which are not invariant under (\ref{trans}).
A similar problem has already been addressed
by Lazarides, Schneider, and S{\o}rensen in \cite{Laza+:93},
where solutions differing by
the sign-alternatig factor and yet having identical energies have been
found in a BCS-like microscopic model of multilayer superconductors,
depending on the sign of the coupling between adjacent layers.
Attractive interaction
leads to positive coupling constant and uniform ground state,
while repulsive interaction results in negative coupling constant and
alternating ground state. The authors
suggest two experimental settings which can determine the negative
sign of
this constant: a superconducting loop able to detect trapped
half-integer flux quanta, and a DC SQUID configuration able to detect
the presence of \(\pi\)-junctions, corresponding to alternating
solutions.\\

\section{Acknowledgements}
This work was supported by the Science Research
Committee (Poland) under grant No. 5 TO7A 040 22.

\thebibliography{13}
\bibliography{}

\bibitem{Abrikosov:book}A.\,A. Abrikosov,
Fundamentals of the Theory of Metals, North--Holland 1988 

\bibitem{Blatter+:94}G. Blatter, M.\,V. Feigel'man, V.\,B. Geshkenbein, A.\,I. Larkin and V.\,M. Vinokur,
{\em Vortices in high--temperature superconductors,} Reviews of Modern Physics {\bf 66}, 1125(1994) 

\bibitem{Rogula:99}D. Rogula,
{\em Dynamics of magnetic flux lines and critical fields in high
T\(_c\) superconductors}, Journal of Technical Physics {\bf 40}
(1999), p.383 

\bibitem{Anderson:92}P.\,W. Anderson,
{\em Intralayer tunneling mechanism for high--T\(_c\)
superconductivity: comparison with c-axis infrared experiments}
included in: P.W.Anderson, The theory of superconductivity in the
high--$T_c$ cuprates, Princeton University Press 1997  

\bibitem{Anderson:97:book}P.\,W. Anderson,
{\em The theory of superconductivity in the high--$T_c$
cuprates,} Princeton University Press 1997  

\bibitem{Lawrence+Doniach:71}W.\,E. Lawrence and S. Doniach,
{\em Theory of layer structure superconductors,} in Proceedings
of the Twelfth International Conference on Low Temperature
Physics, Kyoto, ed. E.Kanda, Academic Press of Japan 1971, p.361 

\bibitem{Theo:88}S.Theodorakis
{\em A phenomenological model for high \(T_c\) supercoductors,}
Physics Lett. A, {\bf 132}, 287(1988)

\bibitem{Laza+:93}N.Lazarides,T.Schneider,M.P.Sorensen,
{\em Mulyilayer high-temperature superconductors,}
Physica C, {\bf 210}, 228(1993)

\bibitem{Theo:93}S.Theodorakis
{\em A new kinetic term in the phenomenology of layered
supercoductors}, Physica C, {\bf 216}, 173(1993)

\bibitem{Ketterson+Song}J.\,B. Ketterson and S.\,N. Song,
Superconductivity,       Cambridge Univ. Press 1999 

\bibitem{Kuple:2001}S. V. Kuplevakhsky,
{\em Exact solutions of the Lawrence-Doniach model in parallel
magnetic fields,}
Phys. Rev. B, {\bf 63}, 054508(2001)

\bibitem{Krasnov:2001}V.\,M. Krasnov,
{\em In-plane fluxon in layered superconductors with arbitrary number
of layers,} Physical Review B {\bf 63}, 064519-11(2001) 

\bibitem{Sztyren:2002}M. Sztyren,
{\em Bi-continuum modelling of layered structures and crystalline
interfaces,}Journal of Technical Physics, {\bf 43}, 265(2006)

\bibitem{Sztyren:2003}M. Sztyren,
{\em Higher grade hybrid model of layered superconductors}, arXiv.org, 
cond-mat/0312673

\bibitem{Rogula+Sztyren:2006}D. Rogula, M. Sztyren,
{\em Long-range Josephson couplings in superconducting systems,}
Journal of Technical Physics, {\bf 47}, (2006)

\bibitem{Born+Huang}M. Born, K. Huang, {\em Dynamical Theory of
Crystal Lattices}, Clarendon Press 1954

\newpage
\vskip 0pt
\noindent \hfil\hbox{\epsffile{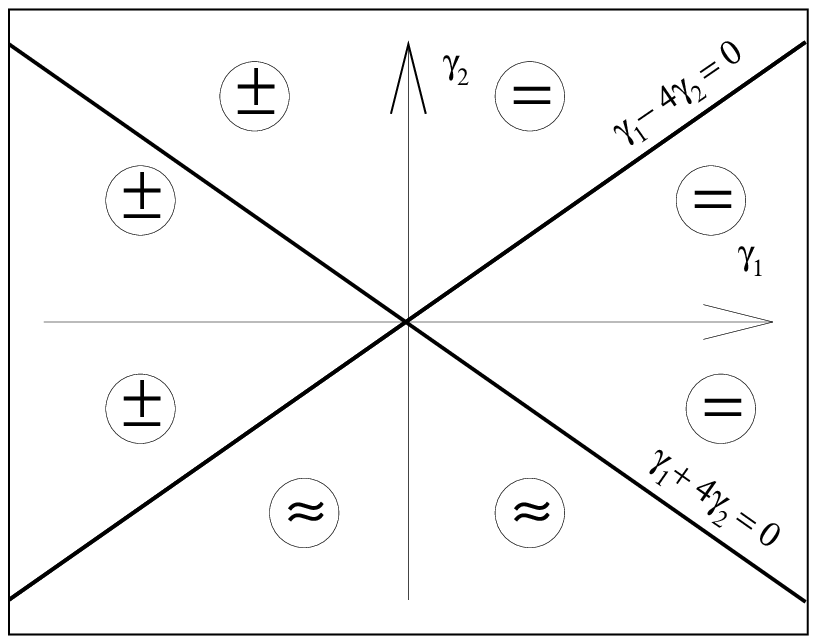}}\hfil\\
\vskip -1.0\baselineskip
\begin{center}
\label{figa}
Figure 1. The regions of stability: uniform
         \(\bigcirc\)\hskip-0.9em\(=\),\\
alternating \(\bigcirc\)\hskip-0.9em\(\pm\),
 and phase modulated
\(\bigcirc\)\hskip-0.9em\(\approx\)
states.

\end{center}


\vskip 0pt
\noindent \hfil\hbox{\epsffile{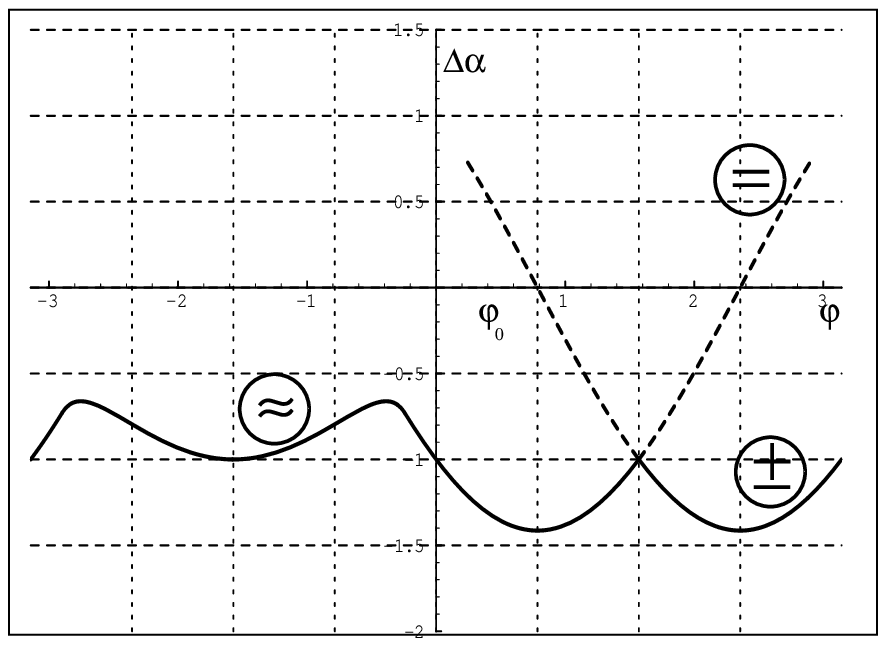}}\hfil\\
\vskip -1.0\baselineskip
\begin{center}
\label{figb}Figure 2. Enhancement of superconductivity\\
by Josephson couplings:
$\Delta\alpha$ vs. $\varphi$.
\end{center}

\end{document}